\documentclass[a4paper,fleqn,usenatbib]{mnras}

\usepackage{newtxtext,newtxmath}
\usepackage[T1]{fontenc}
\usepackage{ae,aecompl}

\usepackage{graphicx}	% Including figure files
\usepackage{times}

\title[The surface abundances of RSGs at core-collapse]{The surface abundances of Red Supergiants at core-collapse}

\author[B.\ Davies \& L.\ Dessart]{
Ben Davies,$^{1}$\thanks{E-mail: b.davies@ljmu.ac.uk}
and Luc Dessart$^{2}$
\\
$^{1}$Astrophysics Research Institute, Liverpool John Moores 
University, Liverpool Science Park ic2, \\ 146 Brownlow Hill, Liverpool, L3 5RF, UK\\
$^{2}$Unidad Mixta Internacional Franco-Chilena de Astronom\'ia (CNRS, UMI 3386), Departamento de Astronom\'ia, Universidad de Chile, \\ Camino El Observatorio 1515, Las Condes, Santiago, Chile}

\date{Accepted XXX. Received YYY; in original form ZZZ}

\pubyear{2018}

\begin{document}
\label{firstpage}
\pagerange{\pageref{firstpage}--\pageref{lastpage}}
\maketitle

\begin{abstract}
In the first weeks-to-months of a type II-P supernova (SN), the spectrum formation region is within the hydrogen-rich envelope of the exploding star. Optical spectra taken within a few days of the SN explosion, when the photosphere is hot, show features of ionised carbon, nitrogen and oxygen, as well as hydrogen and helium. Quantitative analysis of this very early phase may therefore constrain the chemical abundances of the stellar envelope at the point of core-collapse. Using existing and new evolutionary calculations for Red Supergiants (RSGs), we show that the predictions for the terminal surface [C/N] ratio is correlated with the initial mass of the progenitor star. Specifically, a star with an initial mass above 20M$_{\odot}$ exploding in the RSG phase should have an unequivocal signal of a low [C/N] abundance. Furthermore, we show that the model predictions are relatively insensitive to uncertainties in the treatment of convective mixing. Although there is a dependence on initial rotation, this can be dealt with in a probabilistic sense by convolving the model predictions with the observed distribution of stellar rotation rates. Using numerical experiments, we present a strategy for using very early-time spectroscopy to determine the upper limit to the progenitor mass distribution for type II-P SNe. 
\end{abstract}


%%%%%%%%%%%%%%%%%%%%%%%%%%%%%%%%%%%%%%%%%%%%%%%%%%
\def\ga{\mathrel{\hbox{\rlap{\hbox{\lower4pt\hbox{$\sim$}}}\hbox{$>$}}}}
\def\la{\mathrel{\hbox{\rlap{\hbox{\lower4pt\hbox{$\sim$}}}\hbox{$<$}}}}
\def\msunyr{$M_{\odot}{\rm yr}$^{-1}$}
\def\msun{M$_{\odot}$}
\def\zsun{$Z_{\odot}$}
\def\rsun{$R_{\odot}$}
\def\minit{$M_{\rm init}$}
\def\lsun{$L_{\odot}$}
\def\mdot{$\dot{M}$}
\def\mmax{$M_{\rm max}$}
\def\mdotdj{$\dot{M}_{\rm dJ}$}
\def\lbol{$L$\mbox{$_{\rm bol}$}}
\def\kms{\,km~s$^{-1}$}
\def\EW{$W_{\lambda}$}
\def\arcsec{$^{\prime \prime}$}
\def\arcmin{$^{\prime}$}
\def\teff{$T_{\rm eff}$}
\def\Teff{$T_{\rm eff}$}
\def\logg{$\log g$}
\def\logz{$\log Z$}
\def\vdisp{$v_{\rm disp}$}
\def\um{$\mu$m}
\def\chisq{$\chi^{2}$}
\def\AV{$A_{V}$}
\def\hminus{H$^{-}$}
\def\Hminus{H$^{-}$}
\def\ebmv{$E(B-V)$}
\def\mdyn{$M_{\rm dyn}$}
\def\mphot{$M_{\rm phot}$}
\def\cnterm{[C/N]$_{\rm term}$}
\newcommand{\fig}[1]{Fig.\ \ref{#1}}
\newcommand{\Fig}[1]{Figure \ref{#1}}
\def\one{{\,\sc i}}
\def\two{{\,\sc ii}}
\def\three{{\,\sc iii}}
\def\four{{\,\sc iv}}
\def\five{{\sc v}}
\def\six{{\sc vi}}
\def\sev{{\sc vii}}
\def\xii{{\sc xii}}
\def\mesa{{\sc mesa}}
\def\cmfgen{{\sc cmfgen}}

\newcommand{\newt}[1]{ {#1} }

%%%%%%%%%%%%%%%%% BODY OF PAPER %%%%%%%%%%%%%%%%%%

\section{Introduction}
Pre-explosion imaging of the sites of nearby core-collapse supernovae (SNe) has provided unequivocal evidence for the progenitors of hydrogen-rich `plateau' (type II-P) SNe being Red Supergiants (RSGs) \citep{Smartt04,Smartt09}. From these data it is then possible to make quantitative comparisons between the properties of the exploding star and the predictions made by stellar evolution models about the terminal properties of massive stars. 

One such test is to determine the initial masses \minit\ of a sample of II-P progenitors and compare the observed mass range to model predictions. Early results suggested that perhaps the upper mass limit for II-P progenitors was substantially lower than expected, $\sim$17\msun\ as opposed to the $\ga$25\msun\ predicted by models \citep{Smartt09,Smartt15}. However, this result has since been challenged by \citet{Davies-Beasor18}, who argued that systematic errors in the conversion of pre-explosion brightness to initial mass, combined with a small sample size, reduce the statistical significance of this result to well below 3$\sigma$. 

Other authors have attempted to infer the mass of the progenitor by studying the SN radiation \citep[e.g.][]{Bersten11,Dessart13,Utrobin-Chugai17}. Modelling of the light curve and photospheric properties during the plateau phase can yield an estimate of the mass of the H-rich envelope at the point of explosion, though assumptions about the He-core mass and the mass-loss history are required to then obtain \minit. Further, modelling the SN spectrum during the nebular phase has provided estimates of the oxygen yield, which is a  function of the progenitor mass \citep[e.g.][]{Jerkstrand14,Valenti16}. However, as yet, there appears to be a  lack of agreement between the hydro, nebular and pre-explosion estimates for the progenitor initial masses \citep[e.g.][]{Davies-Beasor18}. 

In this paper we present an independent method for inferring the progenitor mass of a II-P which utilises the observed abundance ratios in the very early phases of the SN explosion. The surface composition of the progenitor star at core-collapse should prevail throughout a substantial fraction of the H-rich envelope, depending on how mixed it becomes during the explosion. Here, we have chosen to focus on the spectral appearance at very early times of less than a couple of days. The reason for this is twofold: firstly, at these early times, the spectral formation region is hot (several $\times 10^4$\,K), and so the spectra are dominated by high ionization species such as C{\sc iii}-{\sc iv}, N{\sc iii-iv}, and O{\sc v} \citep{Yaron17,d18_13fs}. From these spectral lines it is possible to estimate the pre-explosion CNO abundances, which as we will show are sensitive to the initial mass of the SN progenitor. Secondly, at such early times we can be confident that explosive mixing has not yet modified the abundances within the spectrum formation region of the supernova.

In Sect.\ \ref{sec:evol} we describe how the surface abundances are modified during a star's evolution, and how it depends on the initial mass of the star. In Sect.\ \ref{sec:spectra} we show the effect of these abundances on the early spectra. In Sect.\ \ref{sec:prospects} we discuss the prospects for using early-time spectra to determine progenitor masses. We conclude in Sect.~\ref{sec:conc}.

\section{Evolutionary models} \label{sec:evol}
In this Section we introduce the chemical signatures of a massive progenitor that can be seen in SN spectra. These signatures relate to the abundances of carbon, nitrogen and oxygen (CNO). Broadly, core hydrogen burning during the main-sequence (MS) alters the relative CNO mix in the centre of the star. As a result of the convective mixing in the centre of the star while on the MS, and the dredge-up by the \newt{surface and intermediate convective zones} as the star becomes a RSG, this polluted material is brought to the surface.

To demonstrate this effect and study its causes, we study two sets of evolutionary models, both computed at Solar metallicity. The first we use are the densely-sampled mass-tracks created using the MESA code \citep{Paxton11,Paxton13,Paxton15} published by \citet[][hereafter the MIST models]{MIST}. We compare the results of these models to a second set of models, those published by \citet{Ekstrom12} using the Geneva evolutionary code (hereafter E12). Both of these works study non-rotating and fast-rotating (40\% of breakup) stars, employ similar prescriptions for mass-loss \mdot\ on the main-sequence and cool supergiant phases\footnote{Subtle differences in the \mdot\ implementation will be discussed later.}, and evolve the stars to the point of carbon depletion in the core. However, there are key differences in the way the two suites of models treat convection and interior mixing. Our reason for choosing particularly these two sets of models is to study the robustness of their predictions to the details of convection. 

The approach of the E12 models is to adopt the Schwarzschild criterion for convective instability. The energy transport in the convective zones is dealt with using mixing length theory (MLT), with a mixing length parameter of $\alpha = 1.6$. Convective overshooting from the core is implemented by extending the mixing boundary beyond the edge of the \newt{convective core by 0.1 times the local pressure scale height}. Since the Ledoux criterion is not used, there is no semi-convection in the E12 models. 

In the MIST models, convective energy transport is again treated using MLT with $\alpha = 1.82$. However, this time the criterion for convective instability comes from Ledoux, which adds in the stabilizing effect of any composition gradient. Regions which are Schwarzschild-unstable but Ledoux-stable are mixed according to a semi-convection parameter of $\alpha_{\rm semi} = 0.1$. Another difference from E12 is the treatment of core overshooting -- this time, the diffusion coefficient falls off exponentially from the edge of the core with a characteristic lengthscale of \newt{0.016 times the local pressure scale-height}. \newt{One further difference to the E12 is that MIST adopts overshoot at the base of the intermediate and surface convective layers (sometimes referred to as `undershoot'), as well as at the top. Both overshoot on both edges of the convective shells is characterized with $\alpha_{\rm os}$=0.0174.}

In the following sections, we use these two suites of models with contrasting treatment of internal mixing to firstly demonstrate the trend of surface CNO abundances with initial mass, to explain the underlying causes of this effect, and its dependence on the details of internal mixing.

\begin{figure}
\begin{center}
\includegraphics[width=8.5cm]{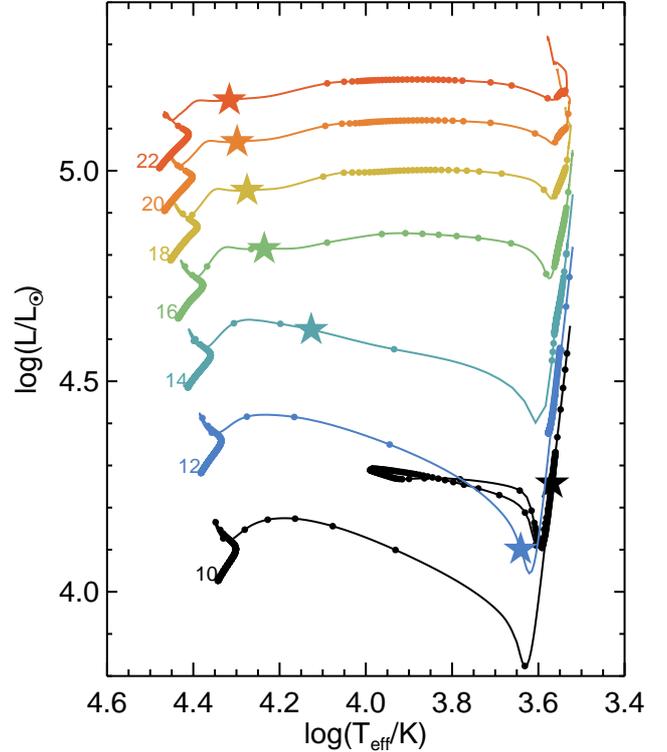}
\caption{Evolutionary tracks of the non-rotating MIST models from midway through the main-sequence to core carbon depletion. The stars indicate the point of core He ignition. The filled circles are the evolutionary tracks resampled onto an evenly-spaced time axis with spacing 5000yrs, to illustrate how long the stars spend in each phase.}
\label{fig:heburn}
\end{center}
\end{figure}

\subsection{Evolution across the H-R diagram} \label{sec:evolution}
Before we look at the trends of surface properties with increasing initial mass, we first look in detail at what happens in a star as it crosses the H-R diagram to the RSG phase. We do this in order to understand how and when dredge-up takes place, since it is this event which is primarily responsible for the surface abundances at the end of the RSG phase. 

Once a star of initial mass $\sim$10\msun\ leaves the MS, it will cross the H-R diagram very rapidly on a timescale comparable to the thermal timescale of the core. Higher mass stars, however, cross the H-R diagram more slowly.  The initial mass at which this change in evolutionary timescale first occurs can be inferred from the morphology of the evolutionary track. In the case of rapid red-ward evolution of the star, the expansion is primarily adiabatic, and the star's luminosity $L$ decreases during this phase. In contrast, slower red-ward evolution is quasi-hydrostatic, and tends to occur at almost constant $L$. For almost all major contemporary single-star evolutionary tracks, the transition from rapid (decreasing $L$) to slow ($\sim$constant $L$) crossing of the yellow void is seen to occur somewhere between 15-20\msun.

There are two factors which affect the crossing of the H-R diagram. These are the point of core He ignition, and the location of the intermediate convective zone with respect to the hydrogen burning shell. Below we explore each of these factors in more detail. To do this, we have taken the MIST `inlist' files from \citet{MIST} and recomputed the evolution\footnote{Our calculations were performed using {\tt mesa} v7503, following \citet{MIST}. } of selected models to obtain the internal profiles as a function of time. For details on these models, including the choices made with regard to mass resolution and timesteps, we refer the reader to \citet{MIST} \citep[see also][]{Farmer16}.

\begin{figure*}
\centering
\includegraphics[width=17cm]{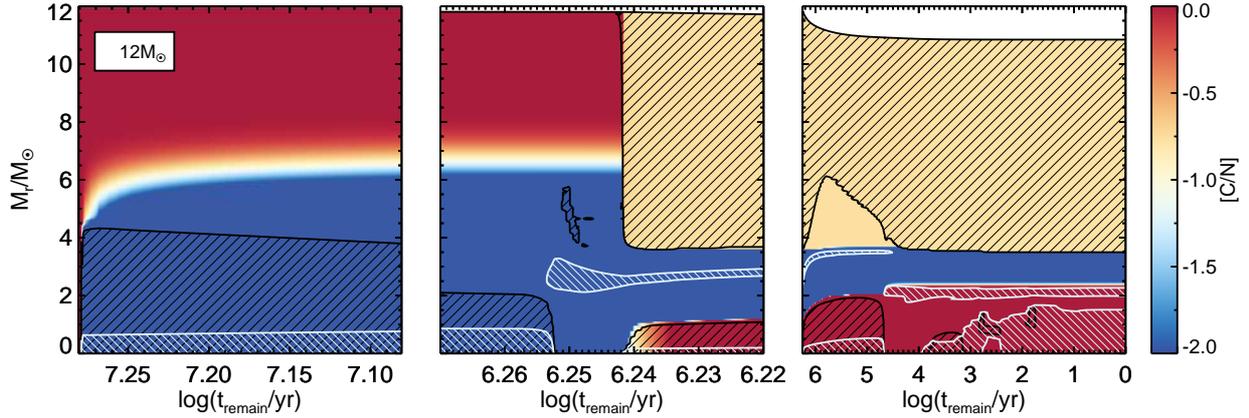}
\caption{Kippenhahn diagram for the MIST 12\msun\ non-rotating model with standard cool-star mass-loss. The colour scale shows the C/N ratio relative to Solar; the black crosshatch areas show the regions of strong convection; and the white crosshatch areas show the regions of strong nuclear burning. The three panels show the main-sequence evolution, a zoom of the end of core H-burning and beginning of core He-burning, and from core He burning to core C depletion. }
\label{fig:12kipp_nonrot}
\end{figure*}

\begin{figure*}
\centering
\includegraphics[width=17cm]{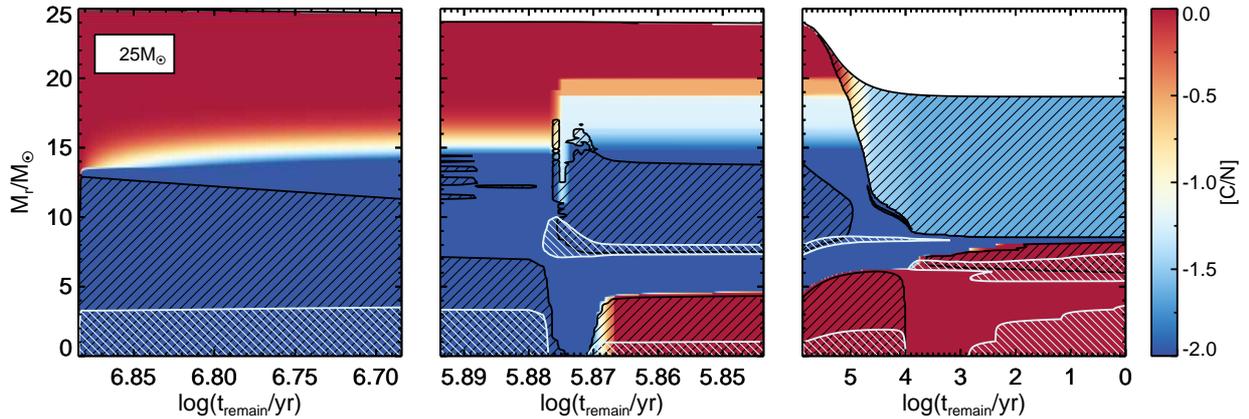}
\caption{Same as Fig. \ref{fig:12kipp_nonrot} but for the 25\msun\ MIST model with no rotation and standard mass-loss.}
\label{fig:25kipp_nonrot}
\end{figure*}

%\subsubsection{Core He ignition}
%The central temperature $T_c$ of a massive main-sequence star scales as $T_c \sim M^x$, where $x \simeq 0.25$ follows from simple homology for a star powered by the CNO cycle. Once H is exhausted, the core contracts and heats up, accompanied by an expansion of the envelope and a traversal of the H-R diagram, until the threshold for triple-$\alpha$ burning is reached at $T_c \simeq 10^8$K. Clearly, more massive stars reach this threshold sooner in their evolution, meaning that core He burning begins at a higher effective temperature \teff\ for stars of higher initial masses. For stars with \minit$\ga$15\msun, He ignites while the star is still close to the MS turn-off. This can help to stall the star's evolution across the H-R diagram, causing it to spend longer in a YSG phase prior to becoming an RSG. This is illustrated in \fig{fig:heburn}. The filled `star' symbols denote the point of core He ignition, while the circles are points in the star's evolution interpolated onto evenly-sampled timesteps of 5000yrs, to illustrate how long stars spend at each phase. The plot shows that for stars with initial masses of 16\msun\ and higher, core He burning begins at \teff$>$16kK, and the star experiences a prolonged YSG phase prior to dredge-up.  

\subsubsection{The intermediate convective zone} 
%One other factor which can affect how a star crosses the H-R diagram is the development of an intermediate
The main factor affecting how a star crosses the H-R diagram is the development of an intermediate convective zone (ICZ) at the end of the MS, above the convective core but well below the surface of the star. The details of the behaviour of the ICZ are explored in Figs.\ \ref{fig:12kipp_nonrot} and \ref{fig:25kipp_nonrot}. In the first of these figures, we present Kippenhahn diagrams of three different stages in the evolution of non-rotating, 12\msun\ star. In the left panel, we see a convective core (black cross-hatched region) that gradually contracts in mass throughout the MS. The centre panel shows the end of the MS, where the end of core H-burning coincides with the formation of the ICZ and the H-burning shell (denoted by the white cross-hatched region), which in this model do not strongly overlap. This contrasts with the 25\msun\ non-rotating model (centre panel of \fig{fig:25kipp_nonrot}), where there is a large degree of overlap between the H-burning shell and the ICZ. These two features first overlap at around 16\msun, and are well separated at lower masses. The overlap ensures that fresh H is continually provided to the shell, increasing the energy generation rate, and helps to slow the evolution across the H-R diagram in the time between core H exhaustion and core He ignition.

%At the end of the MS (centre panel of \fig{fig:15kipp_nonrot}) the core contracts in size, and the H-burning shell develops along with an ICZ. Since the ICZ does not strongly overlap the H-burning shell in this model, the star rapidly crosses the H-R diagram and develops an outer convection zone that reaches from the surface down to the inner 7\msun. This dredges up material from deep within the polluted zone. This results in a rapid decrease in the surface [C/N], illustrated by the change in plotting colour of the envelope. During the later stages of nuclear burning (right panel of \fig{fig:15kipp_nonrot}) the surface convective zone deepens to the inner $\sim$5\msun, further decreasing the surface [C/N].

The depth in the star at which the ICZ forms after leaving the main sequence is perhaps the evolutionary aspect most sensitive to the treatment of convection. This was discussed by \citet[][]{Georgy14}, who experimented with switching between Ledoux and Schwarzschild criteria, and found that the stabilizing effect of the composition gradient lifted the ICZ to higher layers of the star in the Ledoux model compared to the Schwarzschild model. In E12, the default is to use Schwarzschild, whereas in MIST the Ledoux criterion is employed with a semi-convection coefficient of $\alpha_{\rm semi} = 0.1$. The important property of the ICZ in terms of the evolution of the star is whether or not it overlaps the H-burning shell, and this is more likely at higher initial stellar masses.

\subsubsection{The evolution of surface abundances} \label{sec:surfabund}
In Figs.\ \ref{fig:12kipp_nonrot} and \ref{fig:25kipp_nonrot}, the colour scale represents the [C/N] ratio: red represents unpolluted material with the star's initial relative abundances; blue indicates zones heavily polluted by the products of CNO-processing. The left panel focuses on the beginning of the main-sequence (MS). The polluted zone (the region colour-coded as blue in \fig{fig:12kipp_nonrot}) of the 12\msun\ model initially coincides with the boundary of the convective core at $M_r \simeq$4.5\msun. By the end of the main-sequence (centre panel), the core convective zone shrinks in mass to $M_r \simeq 4$\msun, but the polluted zone moves outwards to $M_r \simeq 6$\msun. The latter effect is caused by the C abundance in the radiative zone being higher than that in the outer convective core, meaning that CNO burning at the bottom of the radiative zone can dominate that at the edge of the convective core. Whilst relatively little energy is generated by this burning at the base of the radiative zone, it can still substantially modify the relative CNO abundances.

%In \fig{fig:15kipp_nonrot} we plot the evolution of a 15\msun\ non-rotating star using a Kippenhahn diagram. The plots are colour-coded by the [C/N] ratio: red represents material with the star's initial abundances; blue indicates zones heavily polluted by CNO-processed material. The left panel focuses on the beginning of the main-sequence (MS). The polluted zone (i.e.\ the region colour-coded as blue in \fig{fig:15kipp_nonrot}) initially coincides with the boundary of the convective core at $M_r \simeq$6\msun. By the end of the main-sequence (centre panel), the core convective zone shrinks in mass to $M_r \simeq 3$\msun, but the polluted zone moves outwards to $M_r \simeq 8$\msun. The latter effect is caused by the C abundance in the radiative zone being higher than that in the outer convective core, meaning that CNO burning at the bottom of the radiative zone can dominate that at the edge of the convective core. Whilst relatively little energy is generated by this burning at the base of the radiative zone, it can still substantially modify the relative CNO abundances. %FOR SOME REASON IT WONT COMPILE WITHOUT MORE TEXT HERE

After the 12\msun\ star has left the MS, it rapidly crosses the H-R diagram and becomes an RSG. At this point the star develops an outer convection zone that reaches from the surface down to the inner 4\msun\ which dredges up material from deep within the polluted zone. This results in a rapid decrease in the surface [C/N], illustrated by the change in plotting colour of the envelope. During the later stages of nuclear burning the surface convective zone may deepen, further decreasing the surface [C/N].

The same qualitative behaviour is seen in the 25\msun\ model as in the 12\msun\ model. However, the terminal abundances in the envelope are clearly different, with the more massive star having a lower [C/N] ratio. This is caused primarily by the larger convective core as a fraction of the total mass of the star at ZAMS: the larger the core, the greater the reservoir of polluted material available to be dredged-up at the RSG phase. In the centre panels of Figs.\ \ref{fig:12kipp_nonrot} and \ref{fig:25kipp_nonrot}, we see that the interior polluted zone (coloured blue) of the 12\msun\ star reaches the inner 6\msun\ (50\% of the star by mass), whereas in the 25\msun\ star it's the inner 15\msun\ (60\%). This trend of decreasing [C/N] as a function of increasing initial mass is studied further in Sect.\ \ref{sec:cdivn}.

\subsubsection{The effect of mass-loss on surface abundances}
One further factor which can affect the surface abundances after dredge-up is mass-loss. Specifically, if a large amount of the envelope has been lost {\it prior} to dredge-up, then the mass of the unpolluted material to be mixed in with the polluted material is decreased, resulting in a lower [C/N] ratio post dredge-up. During the MS, the fraction of mass lost by the star is negligible for those initial masses studied here ($\la$1\msun). However, an erroneous result can occur if the star crosses the H-R diagram at a slower pace such that it experiences a prolonged YSG phase prior to dredge-up, which we now describe. 

In most evolutionary codes, mass-loss rates \mdot\ are implemented using analytical formulae which relate \mdot\ to the star's luminosity, temperature and metallicity. Once the star evolved to temperatures below $\sim$10kK, the mass-loss `prescription' is switched to that of \citet{deJager88}. This prescription is a functional fit to empirically-derived mass-loss rates from various studies in the 1970s/80s for a sample of stars spanning spectral types from O to M. The shortcomings of this prescription during the RSG phase have been pointed out by \citet{Beasor-Davies18}, but its description of mass-loss during the YSG phase is perhaps even more uncertain. Here, the de~Jager mass-loss rates are based on only seven stars, many of which are well-known extreme objects such as IRC~+10\,420. Further, the individual \mdot\ measurements are dubious, and the dispersion of the individual stars about the best-fit relation is as large as $\pm$0.7dex. 

Nevertheless, the de~Jager prescription is adopted almost universally by evolutionary codes for massive cool stars. Should a model star experience a prolonged YSG phase, the large \mdot\ assigned to the star by the de~Jager prescription causes a large amount of envelope to be lost prior to dredge-up. As will be discussed in Sect.\ \ref{sec:cdivn}, this can result in anomalously low terminal [C/N] ratios for these stars.

\subsection{The trend of [C/N] with initial stellar mass} \label{sec:cdivn}
\begin{figure}
\centering
\includegraphics[width=8.5cm,bb=00 10 556 415,clip]{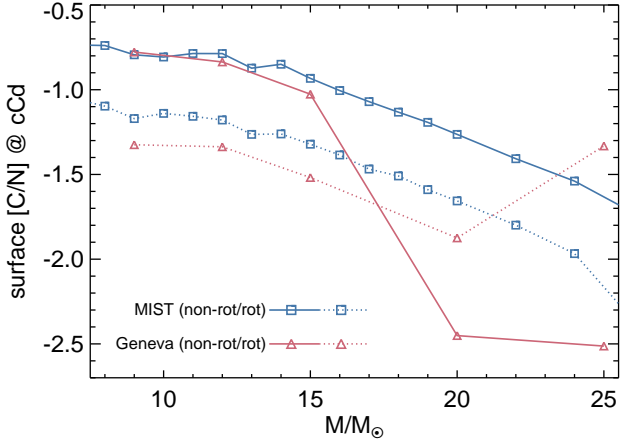}
\caption{The surface ratio of C/N relative to the initial value at the onset of core carbon depletion (cCd) as a function of the star's initial mass, according to the MIST and Geneva (E12) models. Solid lines are non-rotating, dotted lines are rotating at 40\% critical. }
\label{fig:cnterm}
\end{figure}

In \fig{fig:cnterm} we plot the surface C/N ratio relative to Solar at the onset of core carbon depletion (cCd) as a function of initial stellar mass \minit\ for the MIST and E12 models, both non-rotating and rotating. The MIST models, which were computed on a much denser grid than the E12 models, display a clear and monotonic trend of decreasing [C/N] for stars of increasing \minit. As argued in the previous Section, the cause of this trend is primarily the larger convective cores of massive stars at zero-age main-sequence as a fraction of their total mass, which results in a larger (by mass) reservoir of polluted material above the core at the end of the main-sequence. This polluted material is then mixed-in to the convective envelope during the RSG phase. 

Also plotted in  \fig{fig:cnterm} are the MIST rotating models. Increased internal mixing in the rotating stars produces a trend which is qualitatively similar to that of the non-rotating models, but is offset to lower [C/N] by around 0.3-0.4\,dex. This offset is caused by the combination of a larger core, and enhanced internal mixing during the main sequence which allows the products of nuclear burning in the core to diffuse outwards to the intermediate layers where it can be dredged up during the RSG phase. Since the two values of initial rotation present in the MIST models (zero and 40\% or breakup) are somewhat extreme values, we expect a typical star to lie somewhere in between these two values. This then demonstrates that the likely systematic error in [C/N] due to rotation is around $\pm$0.15-0.2\,dex. This is discussed in greater detail in Sect.\ \ref{sec:prospects}.

In terms of the E12 non-rotating/rotating models (plotted in \fig{fig:cnterm} as blue lines), we can see that up to 15\msun\ the two non-rotating tracks produce almost exactly the same behaviour. The rotating E12 models are more depleted in [C/N] by $\sim$0.15dex compared to MIST, due to the different way rotational mixing is treated, but qualitatively the same trend is seen. At 20\msun\ and above, the E12 and MIST results diverge. This is {\it not} caused by the treatment of mixing. In E12, the mass-loss scaling in the \newt{RSG} phase (\teff\ $<$ 5\,kK) has been enhanced by a factor of 3 compared to that in MIST, which contemporary observations suggest is not well justified\footnote{The physical justification in E12 for increasing the mass-loss rates in the cool supergiant phase is that, at masses of 20\msun\ and higher, the star may exceed the classical Eddington limit in their envelope at some point in their post-MS evolution. It is argued that one might expect this to cause \mdot\ to increase. The choice of an enhancement factor of 3 was chosen to mimic the highest \mdot\ RSGs. However, recent studies of RSG mass loss have claimed that total mass lost integrated over the RSG lifetime may be substantially lower than that in models which apply the standard \citep[e.g.][]{deJager88} mass-loss prescriptions \citep{Beasor-Davies18}. }. This three-fold increase in the mass-loss rate, coupled with the much slower rate at which the star crosses the H-R diagram and the high values of \mdot\ assigned to YSGs/RSGs by the \citet{deJager88} prescription, means that a substantial fraction of the stellar envelope can be lost \newt{before the star has reached its minimum \teff}  (see Sect.\ \ref{sec:surfabund}). Hence, \newt{by the time the convective envelope reaches deepest into the star, its mass is greatly reduced compared to a model with the standard mass-loss rate prescription}. This causes the ratio of polluted to pristine material to be larger, and the surface [C/N] to be lower. 

\begin{figure*}
\begin{center}
\includegraphics[width=15cm]{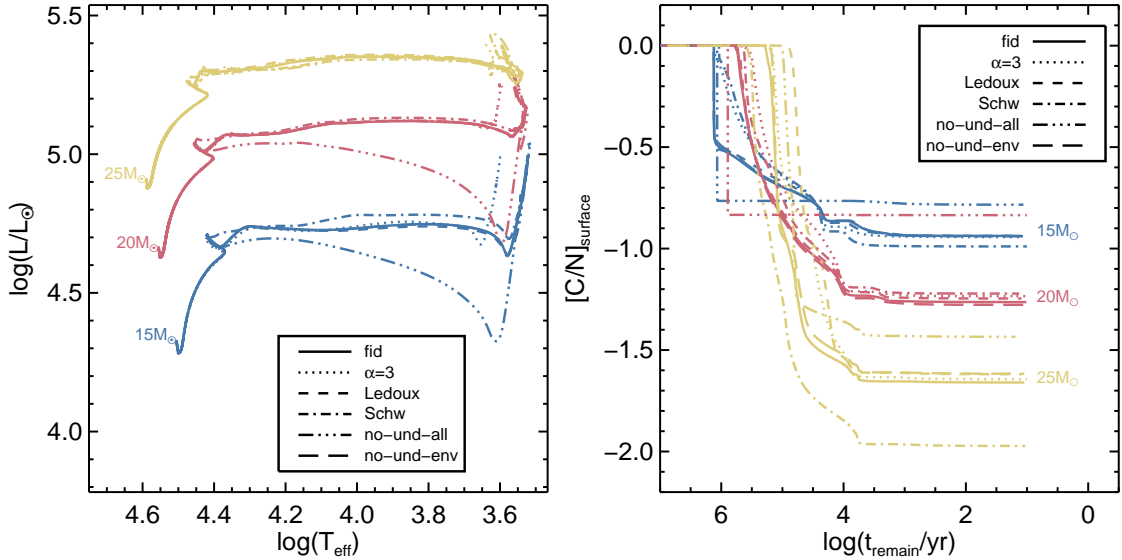}
\caption{The dependence of stellar evolution on the treatment of convection. Left: H-R diagram; right: surface [C/N] as a function of time. See text for details on the families of models. }
\label{fig:details}
\end{center}
\end{figure*}

\subsection{Sensitivity to free parameters relating to convection} \label{sec:conv}
The similarity of the E12 and MIST models up to 15\msun\ hints that the treatment of convection is not so important with regard to the terminal surface abundances \citep[see also][]{Martins-Palacios13}. To perform a more detailed check of this we have computed new models at 15, 20 and 25\msun, in which we have altered how convection is treated. In addition to the fiducial non-rotating MIST models, we have computed further families of models in which the setup is the same as MIST but for one change\footnote{An effect we do not explore here is that of altering the core overshooting parameter. As pointed out by \citet{mey-mae00}, the effects of rotation and core overshoot are largely degenerate, since both serve to mix the outer core with the inner envelope. Since the effects of rotation {\it are} studied here, we do not explore the effects of altering core overshoot explicitly.}:

\begin{itemize}
\item the mixing length parameter is increased to $\alpha_{\rm MLT} = 3.0$. The fiducial value is calibrated against the Sun, however there is no reason to expect that this same value should be employed for all stars of all masses. Increasing the value of $\alpha_{\rm MLT}$ is motivated by the observational evidence that RSGs may have higher \teff\  than in current evolutionary models \citep{rsgteff,Tabernero18}, and that the early colour evolution of SNe II-P require the progenitor RSG to be more compact than for a mixing length characterised by $\alpha_{\rm MLT} = 1.6$ \citep{Dessart13}. The precise value of $\alpha_{\rm MLT} = 3.0$ used here is that tuned by Dessart et al.\ to fit the multi-band lightcurve of SN1999em. Latterly, \citet{Chun18} attempted to tune $\alpha_{\rm MLT}$ by fitting the observed effective temperatures of RSG, finding values of $2.0< \alpha_{\rm MLT} < 2.5$. In this subset of models, the adopted $\alpha_{\rm MLT} = 3.0$ is an extreme limit, which allows us to study the maximal possible effect of this free parameter. 
\item Following the E12 models, we have switched off semi-convection and the Ledoux criterion. These models are referred to as the Schwarzschild models.
\item \newt{In the opposite extreme to the above models, we have computed models which are purely Ledoux (i.e.\ no semi-convection).} 
\item \newt{Finally, we have computed two further sets of models in which we have deactivated overshoot at the base of the convective regions (also known as `undershoot'). In the first of these, we deactivate undershoot at the base of the envelope, which we call `no-und-env'. In the second, we deactivate undershoot entirely, that is no undershoot in the envelope or in any of the intermediate shells (called `no-und-all').  }
\end{itemize}

\newt{
In \fig{fig:details} we plot the evolutionary tracks and the surface abundances as a function of time for the \newt{several families of models (fiducial$\equiv$MIST, $\alpha_{\rm MLT}=3$, Schwarzschild, Ledoux, and the no undershoot models)}. In terms of their paths on the HR diagram, most of the models are very similar. The $\alpha_{\rm MLT}=3$ models have slightly higher \teff\ on the RSG branch, which is expected. A less obvious result is seen in the 15\msun\ and 20\msun\ models with all forms of undershoot deactivated (`no-und-all'). In these models, the lack of undershoot in the ICZ at the end of the MS results in a smaller ICZ, causing less overlap with the H-burning shell. Therefore, these models cross the HRD on a thermal timescale, causing the luminosity to drop as they do so, before rapidly rising again once they arrive on the RSG branch (see Sect.\ \ref{sec:evolution}). 

The pace at which the stars evolve once leaving the MS, and hence the rate at which the surface abundances are polluted, is slightly different in each family of model (right panel of \fig{fig:details}). In the majority of models, the asymptotic [C/N] value at each of the three masses is almost indistinguishable regardless of how convection is treated. The exceptions are discussed below. 

\begin{itemize}
\item In the `no-und-all' models, all masses have lower terminal [C/N] by between 0.2-0.4dex. This is caused by more rapid dredge-up at the end of the MS (in the 15\msun\ and 20\msun\ models), as well as the smaller ICZ reducing the amount of material mixed from deep in the star to the intermediate zones, from where it is dredged to the surface in the RSG phase. 

\item The Schwarzschild model evolves to a slightly lower (-0.3\,dex) [C/N], however this is caused by a prolonged YSG phase, which we have argued here is somewhat spurious. Without the stabilising effect of the $\mu$-gradient term, the ICZ at the end of the MS forms at deeper layers in the star with respect to the fiducial model. This has the effect of stalling the star's traversal of the H-R diagram, holding it in the yellow, and subjecting it to a high mass-loss rate (see Sect.\ \ref{sec:surfabund}). This causes a substantial portion of the envelope to be lost prior to dredge-up, and ultimately in the terminal [C/N] being lower. This is similar to what happens to the 20\msun\ model in E12 (see Sect.\ \ref{sec:evol}). 
\end{itemize}

% is slightly different for all models, which is more clearly illustrated in the right panel of \fig{fig:details}. Once leaving the MS, the rate at which the surface abundances are polluted is slightly different in each family of model. However, in the cases of the 15\msun\ and 20\msun\ models, the asymptotic [C/N] value is almost indistinguishable regardless of how convection is treated. At 25\msun, \newt{four out of five of the models are} again almost identical in terms of their final surface [C/N]. The Schwarzschild model evolves to a slightly lower (-0.3\,dex) [C/N], however this is caused by a prolonged YSG phase, which we have argued here is somewhat spurious. Without the stabilising effect of the $\mu$-gradient term, the ICZ at the end of the MS forms at deeper layers in the star with respect to the fiducial model. This has the effect of stalling the star's traversal of the H-R diagram, holding it in the yellow, and subjecting it to a high mass-loss rate (see Sect.\ \ref{sec:surfabund}). This causes a substantial portion of the envelope to be lost prior to dredge-up, and ultimately in the terminal [C/N] being lower. This is similar to what happens to the 20\msun\ model in E12 (see Sect.\ \ref{sec:evol}). 

From these tests, we conclude that the degree of mixing as a function of evolutionary phase is somewhat sensitive to how convection is implemented\footnote{A caveat to our analysis here is that any change that substantially alters the evolution of the star, such as turning off undershooting, would likely require the other model free parameters (e.g. mixing length, overshooting) to be re-calibrated to retain the agreement with observations. The effect this would have on the systematic errors on [C/N] is not known. }. However, by the time the core is depleted of C (within $\sim10^3$yrs of SN), the majority of our variant models have surface [C/N] ratios consistent to within less than 0.1dex at a given initial mass. The biggest sensitivity is seen when we completely deactivate all forms of undershooting, which shifts the terminal [C/N] higher by 0.2-0.4dex, though the same qualitative trend of [C/N] with initial mass is still seen. At the very worst, our variant models have a difference in terminal [C/N] of $\pm$0.25dex (i.e. the difference between the 25\msun\ `Schwarzchild' and `no-und-all' models). A more important source of systematic uncertainty is stellar rotation, which we discuss further in Sect.\ \ref{sec:prospects}. 

}

\begin{figure*}
\begin{center}
\includegraphics[width=17cm]{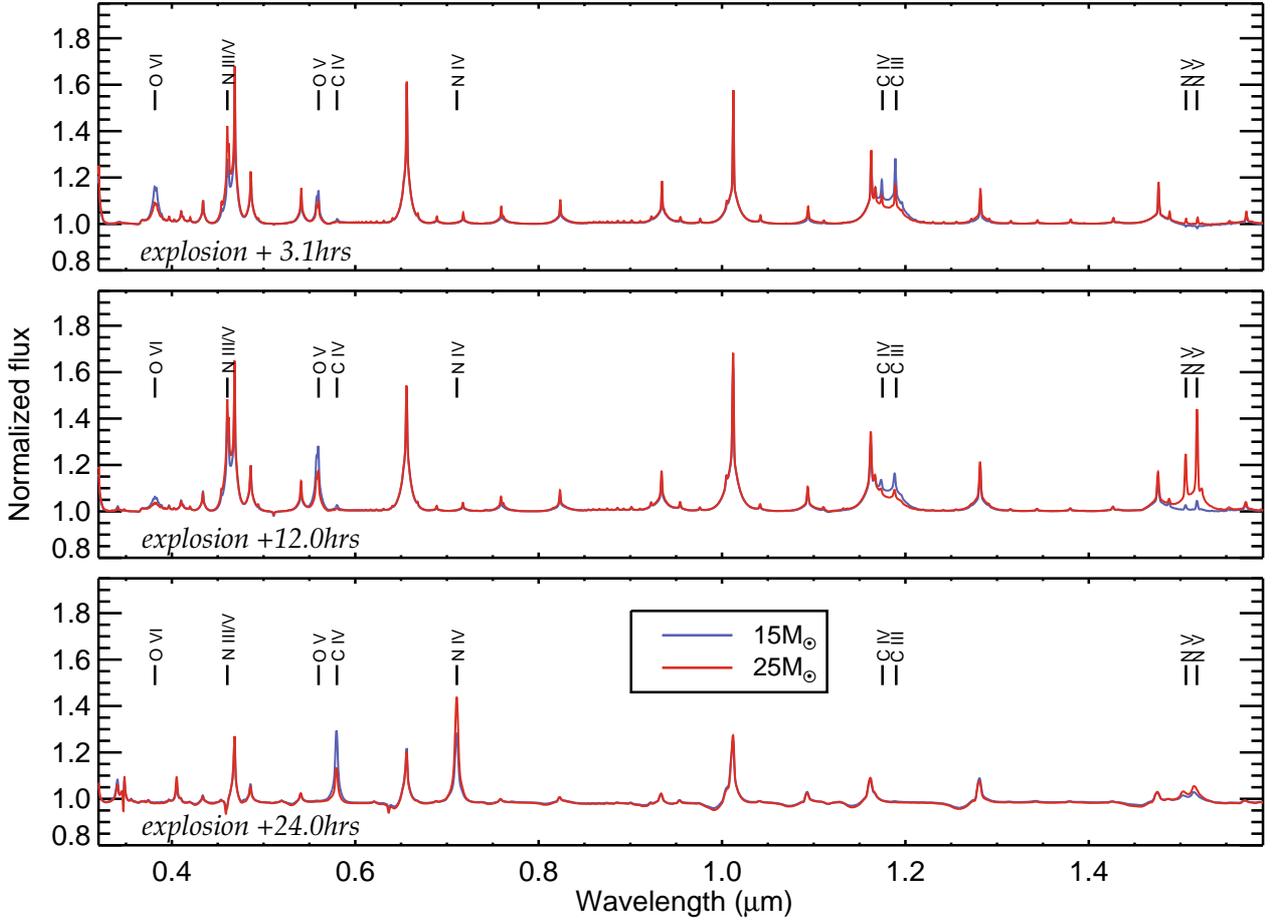}
\caption{Variations in early-time spectra from the MIST non-rotating models at initial masses 15 and 25\msun. The three panels show the model spectra at the time in the bottom left of the plotting panel. Models were generated from the base model r1w5r from \citet{d18_13fs}.}
\label{fig:spec}
\end{center}
\end{figure*}

\section{Expectations for SN spectra} \label{sec:spectra}
In the preceding sections, we have demonstrated that there should be a clear trend of decreasing [C/N] with increasing main sequence mass at the surface of red-supergiant stars. This trend should persist until core collapse and the successful explosion of the star as a Type II SN. In this section, we discuss how CNO abundances may be constrained from Type II SN spectra.

Quantitative spectroscopy of Type II SNe is a very complicated task because of a number of different challenges. The large SN expansion rate produces broad Doppler-broadened line profiles that often suffer from line overlap. This broadening also smears the line flux in velocity (of wavelength) space. It can prevent the clear identification of a given transition or its flux contribution to the line blend. Furthermore, Type II SNe are usually discovered a few days after explosion, at a time when the photosphere has cooled to about 10,000\,K. At that time and beyond, the spectrum tends to show lines of H\one\ and He\one, with only a few lines of CNO elements (typically O\one\,7774\,\AA, which overlaps with a broad atmospheric absorption for SNe at low redshifts; C\one\ lines in the red part of the spectrum). For SN\,1999em, 1999gi, 2005cs or 2006bp, lines of He\two, O\two, or N\two\ were seen as weak features and for a short while \citep{DH05a,baron_05cs_07,quimby_06bp_07,dessart_05cs_06bp}. Abundance determinations require detailed non-LTE radiative transfer modelling. For example, not accounting for departures from LTE poses severe problems with the determination of the He abundance in early time spectra (compare the results from \citealt{eastman_87A_89} with those of \citealt{dh10_87a}). Time dependence in the non-LTE rate equations is also essential for accurate abundance determinations because this time dependence impacts the ionization of all species \citep{UC05,D08_time}. Even with all the necessary physics, the sensitivity of line strengths to abundance changes is weak  (see, e.g., \citealt{DH06}).

Here we exploit the possibility that early time spectra may be obtained for Type II SN with the advent of high-cadence surveys combined with prompt spectroscopic follow-up. In this case, SNe may be more routinely caught when their photosphere is much hotter, exhibiting lines from ionized He and CNO elements. In addition, if the environment of the RSG progenitor is engulfed in a substantial amount of material, these lines will be electron-scattering broadened (rather than Doppler broadened). Because there is no blueshift of the line profile, the peak flux occurs at the rest wavelength and the line identification is facilitated. Several SNe have revealed narrow line profiles of ionized He and CNO  from a few hours to a few days after discovery, for example, SN\,1998S \citep{fassia_98S_01, chugai_98S_01,shivvers_98S_15,D16_2n}, SN\,2013cu \citep{galyam_13cu_14,groh_13cu,grafener_vink_13cu_16}, or SN\,2013fs  \citep{Yaron17}.

For SN\,2013fs, the spectra first exhibit lines of O\five--\six\ and He\two. As the photospheric temperature drops, the ionisation decreases and spectral lines of O\four\ and N\five\ appear, followed after a few days by a dominance of He\one\ and H\one. The rich spectrum in CNO lines during the first hours to days thus turns into one that contains mostly lines of H\one\ and He\one,  which are largely insensitive to the nuclear evolution in the progenitor star.

Using non-LTE radiative transfer and radiation hydrodynamic simulations, \citet{d18_13fs} showed that the early time spectra of SN\,2013fs can be qualitatively reproduced with a fiducial RSG surface composition. Here, we present new simulations for their model r1w5r (performed in the same manner as described in \citealt{d18_13fs}) but this time we adopt the composition given by the surface chemistry of the 15, 20, and 25\,\msun\ models computed with \mesa\ and described in the preceding section. Specifically, we use He, C, N, and O mass fractions listed in Table \ref{tab:snabund}. 

\begin{table}
\caption{Mass fractions of He, C, N and O used in the three models computed here.}
\begin{center}
\begin{tabular}{lcccc}
\hline \hline
$M_{\rm init}$ & X(He) & X(C) & X(N) & X(O) \\
 & & \multicolumn{3}{c}{($\times 10^{-3}$)}  \\
\hline
%15		      & 0.344 & 0.00129 & 0.0033 & 0.0047 \\
%20		      & 0.410 & 0.00087 & 0.0047 & 0.0037 \\
%25		      & 0.474 & 0.00048 & 0.0065 & 0.0022 \\
15		      & 0.344 & 1.29 & 3.3 & 4.7 \\
20		      & 0.410 & 0.87 & 4.7 & 3.7 \\
25		      & 0.474 & 0.48 & 6.5 & 2.2 \\
\hline
\end{tabular}
\end{center}
\label{tab:snabund}
\end{table}%

Figure~\ref{fig:spec} shows the resulting spectra in three different spectral windows at 3.1, 12 and 24\,hr after shock breakout. The lines predicted in the model, which are also present in the observations of SN\,2013fs \citep{Yaron17}, are due to H\,\one\,4340\,\AA, 4862\,\AA, 6562\,\AA,
He\one\,5875\,\AA, 6678\,\AA, He\two\,4686\,\AA, 4860\,\AA,
5411\,\AA, 6562\,\AA, C\four\,5801--5812\,\AA, 7110\,\AA,
N\four\,4057\,\AA, 7122\,\AA, N\five\,4610\,\AA, O\five\,5597\,\AA,
and O\six\,3811--3834\,\AA. In addition, there are strong abundance diagnostics in the near-infrared, with complexes of C\three-\four\ in the $J$-band and N\five\ in the $H$-band. One can clearly see the greater strength of N lines in the higher mass models, and the greater strength of C and O lines in the lower mass models. In particular, the [C/N] ratio is most discernable from the $J$- and $H$-band complexes at 12hr, and in the ratio of C\four\,5801--5812\,\AA\ to N\four\,7122\,\AA\ at 24hrs. In practice, extracting a similar information on CNO abundances from observations will require a set of SNe at early times, so that one can identify a trend in line strengths for similar ionization/temperature conditions in the spectrum formation region.

Finally, we note that analysis of the spectrum of SN2013cu taken at 15.5hrs after explosion \citep{galyam_13cu_14,groh_13cu,grafener_vink_13cu_16} reveals a similarly low [C/N] ratio as that predicted here for massive RSGs. This SN however was a type IIb with a very low H-mass fraction, and had likely lost almost all of its H-rich envelope before exploding. Hence, the abundances seen in the early-time spectrum SN2013cu likely reflect those at the base of the progenitor's H-rich envelope. To lose the entire envelope before exploding would require a very high mass-loss rate which is unlikely to be achieved merely by the wind of a single star \citep{Beasor-Davies18}, and is more likely indicative of mass transfer onto a binary companion. Since binary interaction greatly complicates stellar evolution, the correlation between terminal surface abundances and initial mass would almost certainly be lost. If a substantial H-rich envelope was intact at the point of core-collapse, it would reveal itself as a prolonged `plateau' (or high-brightness) phase in the SN lightcurve, and would indicate that no significant binary interaction had taken place. Therefore, a II-P lightcurve combined with a low [C/N] determined from early-time spectroscopy would together indicate a high mass progenitor. 
 
\begin{figure}
\begin{center}
\includegraphics[width=7.5cm]{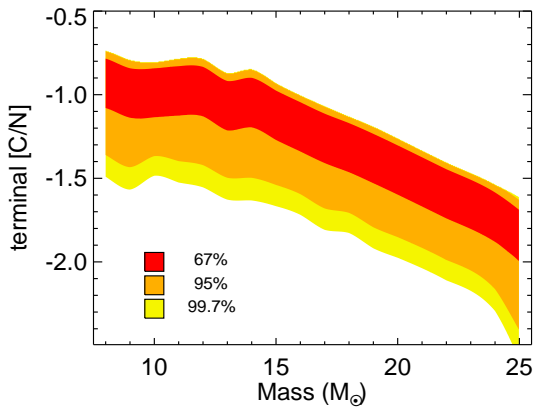}
\caption{The probability distribution of terminal [C/N] ratios as a function of mass factoring in the observed distribution of rotation speeds of massive stars. The red, orange and yellow contours encompass the 67\%, 95\% and 99.7\% probability ranges (analogous to 1, 2 and 3 $\sigma$).}
\label{fig:contour}
\end{center}
\end{figure}

\begin{figure}
\begin{center}
\includegraphics[width=7.5cm]{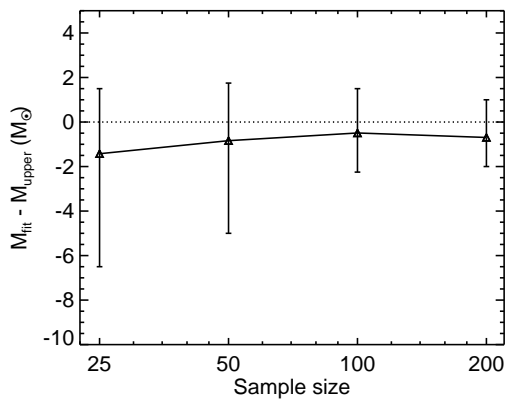}
\caption{The accuracy and precision to which the upper mass limit \mmax\ can be recovered from the terminal [C/N] ratios of a sample of SNe, as a function of sample size. }
\label{fig:MC}
\end{center}
\end{figure}

\section{Prospects} \label{sec:prospects}
Having shown above that the mass-dependence of the RSG surface abundances should be visible in early-time spectra, we now study the prospects of using these observational signatures to determine the mass-function of SN II-P progenitors. We do this by simulating the predicted cumulative distribution of [C/N] ratios for a finite sample of SNe, factoring in both model and observation errors. 

The first step is to define the predicted trend of terminal [C/N] with initial mass, accounting for the model uncertainties. As discussed in Sect.\ \ref{sec:conv}, by far the largest source of uncertainty is that of stellar rotation. Specifically, a model with an initial rotation speed of 40\% of breakup can have a final surface [C/N] ratio 0.4dex lower than a non-rotating star. However, studies of stellar rotation have shown that the distribution of rotation speeds of massive stars is such that the majority have rotation rates halfway between these values \citep{Ramirez-Agudelo13}. By taking the observed rotation rate distribution of Ramirez-Agudelo et al.\footnote{We note that the distribution of rotation rates in \citet{Ramirez-Agudelo13} is that {\it observed} for main-sequence O~stars, rather than that at zero-age main-sequence.}, and making the assumption that the trend of terminal surface [C/N] scales linearly with initial rotation speed\footnote{The knowledge of how terminal [C/N] scales with rotation speed at a fixed initial mass requires the computation of more models, which is beyond the scope of this current work. Here, we make the most simple assumption of a linear scaling between the two.}, we can make a probability density map for the [C/N] ratio as a function of initial mass. This is shown in \fig{fig:contour}, where the red, orange and yellow contours illustrate where we expect 67\%, 95\% and 99.7\% of stars to lie at a given initial mass\footnote{\newt{One major caveat to this experiment pertains to the model predictions of rotational mixing as a function of rotation rate. In both the MIST and E12 models, validation of the degree of rotational mixing is provided by comparing  predictions of surface N abundances for main-sequence massive stars with observations of OB stars. Agreement between the observations and the `rotating' models is interpreted as a satisfactory calibration of the diffusion coefficients for rotational mixing. This is despite the fact that the `rotating' models have rotation rates that are extreme in comparison to those typically observed. Since these diffusion coefficients are largely free parameters, in principle one could achieve a similar agreement between observed and predicted surface N abundances at lower rotation rates. Therefore, the absolute level of surface enrichment as a function of initial rotation rate may not be a robust model prediction.}}.

The next step is to simulate a population of massive stars by randomly sampling numbers from a Salpeter initial mass function (IMF) with an upper mass cutoff \mmax, which we leave as a free parameter. We then use \fig{fig:contour} to randomly assign each star a terminal [C/N] ratio based on the probability distribution shown in this figure. Next, we factor in observational uncertainties by multiplying each star's terminal [C/N] by a Gaussian function centred on unity with a width $\sigma=0.2$, to simulate an experimental uncertainty of $\pm$0.2\,dex\footnote{Though we do not yet know what the experimental uncertainties will be, $\pm$0.2dex is a standard conservative error on abundance work in the modelling of hot star winds. The retrieval of abundances from early-time SN spectra, and degeneracies with other model parameters such as H/He and temperature, will be the subject of a future paper.}. From these steps, we can then determine the expected distribution of [C/N] ratios for a given sample size and \mmax. 

Finally, we take the simulated [C/N] distribution and attempt to recover the input value of \mmax. We do this by first determining the cumulative distribution of the simulated [C/N] ratios. We then compare this to a grid of  similar distributions of the same sample size $\mathcal{N}$ but where \mmax\ is allowed to vary. For a given trial, we measure the most likely \mmax\ from a $\chi^2$-minimization procedure. Since there are large stochastic effects, we repeat the measurement at each \mmax\ and $\mathcal{N}$ $10^4$ times, and determine the posterior probability distributions. 

In \fig{fig:MC} we show the results of this numerical experiment. The plot shows the difference between the input and output \mmax\ and their 67\% probability limits as a function of sample size $\mathcal{N}$. There is a small systematic error in this type of analysis which causes one to underestimate the upper limit to a random distribution of numbers, since one cannot sample from numbers above this limit. However, it is possible to characterize and estimate this systematic error \citep[see also ][]{Davies-Beasor18}. The plot also shows that, for sample sizes greater than 50, the experimental errors on \mmax\ begin to shrink to around $\pm$15\%. Though the current sample size of SNe with early-time spectroscopy is an order of magnitude smaller than this, the current collection rate of $\sim$4-5 per year is already greater than that for SNe with pre-explosion imaging of the progenitors. Since this collection rate continues to grow, it is likely that a sample of $\sim$100 SNe with early-time spectra will be achieved within the next decade.

\section{Conclusions} \label{sec:conc}
\newt{
Using existing and new stellar evolution calculations, we have shown that the surface CNO abundances of Red Supergiants at the end of their lives are correlated with their initial masses. Specifically, the ratio of carbon to nitrogen at the point of core carbon depletion [C/N], which is a diagnostic of how much nuclear processed material has been mixed into the envelope, is shown to decrease with increasing initial mass. We explored the sensitivity of this trend to how convection is treated using families of `variant' models in which we alter semi-convection, mixing length and undershooting. In the majority of cases the effect on [C/N] is less than $\pm$0.1dex, and is at worst $\pm$0.25dex. The dominant source of systematic error is the dependence on initial stellar rotation. }

We have taken the end-points of the evolution calculations and simulated how their supernova (SN) spectra would look at very early times. Though subtle, there is a clear trend of increasing nitrogen and decreasing carbon line strengths at higher initial stellar masses. We argue that in principle this signature can be used to diagnose the initial mass of the SN progenitor. 

Using a simple numerical experiment, we have set out an observing strategy to use the trend of terminal [C/N] with stellar mass to determine the upper limit to the mass range of SN II-P progenitors ($M_{\rm max}$). Assuming that [C/N] can be recovered to precision of $\pm$0.2\,dex, and assuming that the initial rotation rate distribution of stars follows that of Galactic O stars, we show that $M_{\rm max}$ can be determined to within $\pm$3-4\msun\ from a sample size of 50, and to within $\pm$2\msun\ for a sample size of 100.

\section*{Acknowledgements}
We thank the referee Cyril Georgy for the comments and suggestions that helped us improve the paper. For numerous constructive discussions we thank Dave Arnett, Emma Beasor, Sylvia Ekstrom, John Hillier, Raphael Hirschi, Georges Meynet and Nathan Smith. This work made use of the IDL astronomy library, available at {\tt https://idlastro.gsfc.nasa.gov}, and the Coyote IDL graphics library. 

%%%%%%%%%%%%%%%%%%%%%%%%%%%%%%%%%%%%%%%%%%%%%%%%%%



\bibliographystyle{mnras}
%\bibliography{/Users/astbdavi/Google_Drive/drafts/biblio} % if your bibtex file is called example.bib
\bibliography{../../bib_merge_luc} % if your bibtex file is called example.bib


%%%%%%%%%%%%%%%%% APPENDICES %%%%%%%%%%%%%%%%%%%%%

%\appendix


%%%%%%%%%%%%%%%%%%%%%%%%%%%%%%%%%%%%%%%%%%%%%%%%%%

% Don't change these lines
\bsp	% typesetting comment
\label{lastpage}
\end{document}